\documentclass[twocolumn,superscriptaddress,prb]{revtex4-1}
\usepackage{color}
\usepackage{graphicx}
\usepackage{epstopdf}
\usepackage{amsmath}
\usepackage{mathtools}
\usepackage[colorlinks,linkcolor=blue,anchorcolor=blue,citecolor=blue,urlcolor=blue]{hyperref}

\begin{document}
\title{Slater and Mott insulating states in the SU(6) Hubbard model}
\author{Da Wang} 
\affiliation{National Laboratory of Solid State Microstructures and School of Physics, Nanjing University, Nanjing 210093, China}
\author{Lei Wang} 
\affiliation{Beijing National Lab for Condensed Matter Physics and Institute of Physics, Chinese Academy of Sciences, Beijing 100190, China}
\author{Congjun Wu} 
\affiliation{Department of Physics, University of California, San Diego, California 92093, USA}
\begin{abstract}
We perform large scale projector determinant quantum Monte-Carlo simulations
to study the insulating states of the half-filled SU(6) Hubbard model on the square lattice.
The transition from the antiferromagnetic state to the valence bond 
solid state occurs as increasing the Hubbard $U$.
In contrast, in the SU(2) and SU(4) cases antiferromagnetism persists
throughout the entire interaction range.
In the SU(6) case, antiferromagnetism starts to develop in the weak
interacting regime based on the Slater mechanism of Fermi surface nesting.
As $U$ passes a crossover value $U^*/t\approx 9$, the single-particle gap
scales linearly with $U$, marking the onset of Mott physics.
In the Mott regime, antiferromagnetism becomes to be suppressed
as $U$ increases, and vanishes after $U$ passes the critical value $U_{\rm AF,c}/t=13.3\pm 0.05$.
The critical exponents are obtained via critical scalings as
$\nu_{\rm AF}=0.60\pm 0.02$ and $\eta_{\rm AF}=0.44\pm 0.03$.
As $U$ further increases, the valence bond solid ordering appears
exhibiting the anomalous dimension $\eta_{\rm VBS}=0.98\pm 0.01$.
\end{abstract}
\maketitle

\section{introduction}
How repulsive interactions turn a partially filled electron band into the
insulating state is an important question of strong correlation physics.
In the presence of nested Fermi surfaces, the antiferromagnetic (AF) order
appears at infinitesimal interactions based on Fermi surface nesting.
The single-particle gap is at the same order of the AF gap function \cite{Slater_PR_1951}.
Such a state is known as the Slater insulator exhibiting strong charge fluctuations.
On the other hand, charge fluctuations are frozen in the strong interaction
regime due to the large charge gap linearly scaling with the 
repulsive interaction \cite{Mott_PPSA_1949}, and such a state is Mott insulating.
The low energy physics lies in the spin channel arising from the
superexchange among local spin moments.
Mott insulators can even exhibit no symmetry breaking, for example,
the 1D Hubbard model at half-filling exhibits power-law AF correlations
and charge gap, but without long-range AF ordering.

However, in strongly correlated electron systems, the above two pictures
of insulators are often blended together \cite{Imada_RMP_1998,Lee_RMP_2006}.
For example, in the half-filled SU(2) Hubbard model on the square lattice,
Fermi surface nesting leads to the Slater AF state at weak $U$, while the
strong $U$ side is effectively described by the Heisenberg model
and attributed to the Mott AF insulator.
Both regimes exhibit the commensurate Neel ordering smoothly
connected by a crossover \cite{Hirsch_PRB_1985,*Hirsch_PRL_1989,Pruschke_JPCM_2003}.

In recent years, the rapid development of ultra-cold atom physics
provides a new route to investigate strong correlation physics.
It was proposed that cold fermions with multiple spin components
are ideal systems to study high symmetries that typically are met
in high energy physics \cite{Wu_PRL_2003,*Wu_MPLB_2006,Honerkamp_PRL_2004, Wu_2012}.
For example, the spin-3/2 fermion systems possess a generic Sp(4)
symmetry without fine-tuning, which is further enlarged to
 SU(4) when the interaction is spin-independent.
These symmetries play an important role to study novel quantum magnetism
beyond SU(2)
\cite{Wu_PRL_2003,*Wu_MPLB_2006,Wu_2005,Hung_2011,Xu_2008,Wang_PRL_2014,
  Chen_2005}.
The study of high symmetric ultra-cold fermions has been attracting
considerable interests both experimental and theoretical recently
\cite{Wu_PRL_2003,*Wu_MPLB_2006,Honerkamp_PRL_2004,DeSalvo_PRL_2010,
  *Taie_PRL_2010,Krauser_NP_2012,*Taie_NP_2012,*Zhang_S_2014,
  Cazalilla_RPP_2014,*Laflamme_AP_2016}.

As for the SU(4) Hubbard model at half-filling on the square lattice, 
i.e., two fermions per site, a pervious determinant quantum Monte 
Carlo shows that the AF order is non-monotonic as $U$ increases:
After reaching a maximal at $U/t\approx 8$, the AF order starts to
decrease but remains finite throughout the interaction range
simulated $U/t\le 20$ \cite{Wang_PRL_2014}.
Meanwhile, the system exhibits no valence bond solid (VBS)
ordering.
A recent study directly on the SU(4) Heisenberg model with the
one column self-conjugate representation also shows the survival
of the AF order \cite{Kim2019}, hence, the AF order should
also persist through the entire interaction range.
The SU(4) and SU(6) Hubbard models of Dirac fermions
in the honeycomb lattice and the $\pi$-flux square lattice
exhibit the transition from the Dirac semi-metal phase to
VBS state and show the absence of the AF order \cite{Zhou_PRB_2016,Zhou_2018}.
In contrast, the half-filled SU(6) Hubbard model in the square
lattice behaves very differently.
The QMC simulations show a transition from the AF state at weak $U$
to the VBS state at strong $U$ \cite{Wang_PRL_2014,Lang_ppt}.
In the weak $U$ limit, the AF is a direct consequence of the Fermi
surface nesting and the Van Hove singularity, while the VBS
state is a manifestation of the Mott physics.
How such a Slater to Mott transition occurs is an interesting and
open question, which is the main aim of the present work.

On the other hand, the quantum phase transition from the AF state to
the VBS one is argued to be continuous as a result of the
deconfined criticality beyond the Landau-Ginzburg-Wilson paradigm \cite{Senthil_S_2004,*Senthil_PRB_2004,*Levin_PRB_2004}.
Such a prediction has been supported by numerical simulations
in recent years. \cite{Sandvik_PRL_2007,Melko_PRL_2008,*Sandvik_PRL_2010,
  *Kaul_PRL_2012,*Pujari_PRL_2013,*Shao_S_2016,*Nahum_PRX_2015,
  *Wang_PRB_2016,*Assaad_PRX_2016}.
But there are also works claiming the first order phase
transition \cite{Kragset_PRL_2006,*Kuklov_AP_2006,*Kuklov_PRL_2008,
  *Sen_PRB_2010,*Papanikolaou_PRL_2010,*Block_PRL_2013,
  *DEmidio_PRB_2016,*DEmidio_PRL_2017}.
Nearly all these models studied so far are based on quantum spin 
models in which charge fluctuations are frozen.
It would be interesting to directly investigate the transition
between the AF and VBS states based on the fermionic Hubbard model,
which takes into account both charge and spin fluctuations.

An additional motivation of this work is the AF order studied
below belongs to the self-conjugate representation of SU(N),
which could be described by the $U(N)/[U(N/2)\otimes U(N/2)]$
nonlinear $\sigma$-model \cite{MacFarlane_PLB_1979,Duerksen_PRD_1981,Read_PRL_1989,
  *Read_NPB_1989,*Read_PRB_1990}.
The symmetry class is different from the widely studied
CP$^{N-1}$ model which respects the fundamental representation
corresponding to the case of $U(N)/[U(1)\otimes U(N-1)]$.
Therefore, the AF phase transition here (if continuous) belongs
to a different universality class, whose critical exponents
would be desired to calculate to characterize such
a university class.

\begin{figure}
  \includegraphics[width=0.5\textwidth]{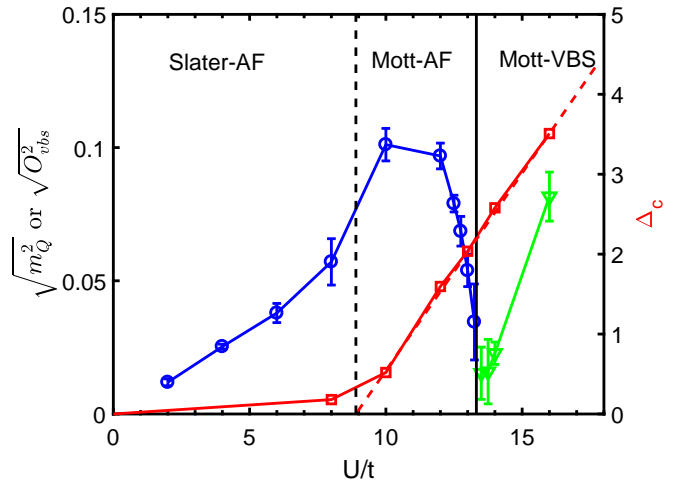}
  \caption{\label{fig:phasediagram}
    Phase diagram of the half-filled SU(6) Hubbard model.
    The AF and VBS order parameters are defined as
    $\sqrt{m_Q^2}$ and $\sqrt{O_{vbs}^2}$ at $L\to \infty$
    marked by the blue and red lines, respectively.
    The error bars are determined by the $95\%$ confidence bounds of
    the least square fittings of the finite size data.
    The single-particle gap $\Delta_c$ at $U/t\ge10$ marked by
    the red line is extracted from the single-particle Green's function
    at $k=(\pi,0)$ and $L=14$.
    The solid black line indicates the transition from AF to VBS
    which is obtained from the data crossing as shown
    in Fig.~\ref{fig:correlationlength_af}.
    The dashed black line is obtained from the linear extrapolation
    (dashed red line) of $\Delta_c$ at large $U$, which indicates
    the crossover from the Slate and Mott regimes.
    (The order parameter values at $U/t\le8$ are taken
    from Ref.~\onlinecite{Wang_PRL_2014}, and
    $\Delta_c$ at $U/t=8$ is taken from Ref.~\onlinecite{Cai_PRB_2013}.)
  }
\end{figure}

In this work, we apply the projector determinant QMC free of the sign problem
to study the half-filled SU(6) Hubbard model on the square lattice.
Our main results are shown in Fig.~\ref{fig:phasediagram}.
From the slope of the single-particle gap, a crossover from the Slater to
Mott insulating regime around $U^*/t\approx 9$ accompanied by the 
obvious enhancement of the AF order.
The AF order reaches the maximum around $U/t\approx 10$ and then starts
to drop as $U$ further increases.
In the Mott insulator side, the vanishing of the AF order occurs
at $U_{\rm AF,c}/t=13.3 \pm 0.05$, and simulations show a continuous transition
with the critical exponents $\nu_{\rm AF}=0.60\pm 0.02$ and
$\eta_{\rm AF}=0.44\pm 0.03$.
As for the appearance of the VBS order, $\eta_{\rm VBS}=0.98\pm0.01$, 
and more nature of this transition will be deferred for a future study.

\section{The model definition and QMC parameters}

The SU($N$) Hubbard model on the square lattice at half-filling
is defined as
\begin{eqnarray}
H=-t\sum_{\langle i,j\rangle,\alpha}\left(c_{i\alpha}^\dag c_{j\alpha}
+h.c.\right)+\frac{U}{2}\sum_{i}\left(n_i-\frac{N}{2}\right)^2, ~~
\label{eq:hamilton}
\end{eqnarray}
where $c_{i\alpha}$ is a fermion annihilation operator with $i$ the site
index and $\alpha$ the flavor index satisfying $1\le\alpha\le N$.
The $t$-term represents hoppings between the nearest neighbour sites,
and $t$ is the hopping integral.
The $U$-term describes the on-site Hubbard interaction
as usual and the onsite particle number
$n_i=\sum_{\alpha=1}^{N} c_{i\alpha}^\dag c_{i\alpha}$.
Eq. \ref{eq:hamilton} satisfies the particle-hole symmetry, hence,
the average particle number per site is fixed at $N/2$, and the
chemical potential $\mu$ is not shown explicitly.
When $N=2$, Eq. \ref{eq:hamilton} goes back to the usual spin-$\frac{1}{2}$
Hubbard model.
In this article, we focus on $N=6$.

The half-filled SU($N$) (with even $N$) Hubbard model on a bipartite
lattice is free of sign problem in auxiliary field QMC simulations as
a result of the particle-hole symmetry\cite{Cai_PRL_2013}, which enables
us to perform large scale simulations.
Details of the algorithm can be found elsewhere
\cite{Assaad_CMP_2008,Wang_PRL_2014,Zhou_PRB_2016}, and will not be
repeated here.
In our simulations, the projection time $\beta=2L$ is used, which
is long enough to achieve convergence for a given linear lattice
size $L$ up to $24$.
The discrete time slice $\Delta\tau=0.05$ is chosen.
For each group of parameters, the simulation is performed on 24 cores
with 1000 Monte Carlo steps for warming up and no less than
1000 steps for measurements on each core.
The exception is the case of $L=24$ which is performed on 48 cores
with 500 Monte-Carlo steps for warming up and no less
then 500 steps for measurements.

For later convenience, we define the following correlation functions.
For the AF order, due to the SU(6) symmetry, we take the diagonal
component of the spin-moment operator
\begin{eqnarray}
m_r=\frac{1}{6}\left( \sum_{\alpha=1}^{3}n_{r\alpha}
-\sum_{\alpha=4}^{6}n_{r\alpha} \right),
\end{eqnarray}
whose largest eigenvalue is normalized to 1/2 as in the case of SU(2).
The Fourier component of spin-moment at $Q=(\pi,\pi)$ corresponds to 
the AF order parameter $m=\frac{1}{L^2}\sum_{r}m_r(-1)^r$.
Due to the finite sizes of QMC simulations, there is no spontaneous symmetry
breaking, and we measure spin structural factor defined as
the equal-time spin-spin correlation,
\begin{eqnarray}
S_{mQ}=\frac{1}{L^2}\sum_{rr'}\langle m_rm_{r'}\rangle (-1)^{r-r'}.
\end{eqnarray}
To describe the VBS order, we define the kinetic dimer operator,
\begin{eqnarray}
d_{r\hat{e}}=\sum_{\alpha=1}^{6}\left(c_{r\alpha}^\dag c_{r+\hat{e}\alpha}
+ c_{r+\hat{e}}^\dag c_{r\alpha} \right),
\end{eqnarray}
where $\hat e=\hat x,\hat y$.
Then the VBS order parameters $d_{\hat x}$ and $d_{\hat y}$ are defined as
the Fourier components at $(\pi,0)$ and $(0,\pi)$, respectively,
\begin{eqnarray}
d_{\hat x (\hat y) }=\frac{1}{L^2}\sum_r d_{r\hat{x} (\hat y)}(-1)^{r_x (r_y)}.
\end{eqnarray}
Again, we directly measure the structure factor of dimer-dimer correlation
\begin{eqnarray}
S_{vbs} =\frac{1}{L^2}\sum_{rr',\hat e}
\langle d_{r\hat{e}}d_{r^\prime \hat{e}}\rangle
(-1)^{r_{\hat e}-r^\prime_{\hat e}}.
\end{eqnarray}
In large $U$ limit (Heisenberg limit), the kinetic dimer order is equivalent to the spin-Peierls
VBS defined as
\begin{eqnarray}
d_{r\hat{e}}\propto \frac{t}{U} \sum_{\alpha\beta}c_{r\alpha}^\dag c_{r\beta} c_{r+\hat{e}\beta}^\dag c_{r+\hat{e}\alpha},
\end{eqnarray}
for the SU($N$) Heisenberg models \cite{Read_PRL_1989,*Read_NPB_1989,*Read_PRB_1990,Marston_PRB_1989}
through the 2nd order perturbation theory. 
(For finite $U$, charge fluctuations may cause the inequivalence of these two kinds of VBS definitions.)
Based on the AF and VBS structure factors $S_{mQ}$ and $S_{vbs}$, we further denote
\begin{eqnarray}\label{eq:order}
m^2_Q(L)  = S_{mQ}/L^2, \ \ \,
O_{vbs}^2 (L) =S_{vbs}/L^2.
\end{eqnarray}
In the presence of long-range ordering of AF and VBS,
$m^2_Q$ and $O_{vbs}^2$ exhibit non-vanishing values in the
thermodynamic limit $L\to \infty$, respectively.

\section{QMC simulation results}

\begin{figure}
\includegraphics[width=0.48\textwidth]{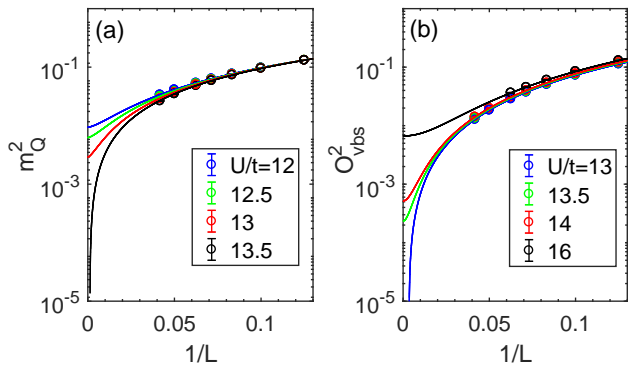}
\caption{\label{fig:extrapolation} Finite size extrapolations
of ($a$) $m^2_Q$  and ($b$) $O_{vbs}^2$
versus $1/L$ at various values of $U$ near the quantum
phase transition point.
The fitting uses the power-law relation given in Eq.~\ref{eq:fitting}.
The logarithmic coordinates are used for the vertical axes
for the order parameter squares.
}
\end{figure}

We first present the single particle gap $\Delta_c$ extracted from
the slope of $\ln G(\tau,k)$, where $G(\tau,k)$ is the single particle
Green's function
defined as $G(\tau,k)=-\frac{1}{L^2}\sum_{rr'}\langle \mathcal{T}_\tau
c_{r\alpha}(\tau) c_{r'\alpha}^\dag \rangle \mathrm{e}^{ik\cdot(r-r')}$.
The momentum $k$ is taken at $(\pi,0)$ on the Fermi surface.
The results for $L=14$ are shown in Fig.~\ref{fig:phasediagram}.
When $\Delta_c \gtrsim1$, it shows very little size dependence,
because it describes the local charge fluctuations with a very
short charge coherence length estimated as
$\xi_c \sim t/\Delta_c\lesssim1$.
Hence, the results at $L=14$ already can be taken as the thermodynamic limit.
An interesting observation is the nearly linear dependence of $\Delta_c$
on $U$ at $U>U^*\approx 9t$, whose slope is very close to 1/2,
indicating the characteristic feature of the Mott insulator.
This is consistent with that in the atomic limit, {\it i.e.},
$t/U\rightarrow 0$, which is simply $U/2$,
the energy cost by adding or removing an electron on the half-filled
Mott insulating background.
On the other hand, in the regime $U<U^*$, $\Delta_c$ keeps at very
small values, which is consistent with the AF insulators based on
the Fermi surface nesting as in a Slater insulator.
Therefore, we take $U^*$ as a crossover from the Slater to Mott
regimes since no symmetry breaking occurs.

Next we consider the AF and VBS orderings near the AF-VBS transition
by performing the finite size extrapolation of $m^2_Q$ and
$O_{vbs}^2$ at $L\to \infty$.
Without a precise knowledge of finite size effects in prior,
we have tried different fitting functions.
It turns out that the usual polynomial (neither square nor cubic) functions
of $1/L$ used in Ref. [\onlinecite{Neuberger_PRB_1989,Sandvik_PRB_1997}]
fail to fit the data.
Instead, a simple (non-integer) power law function
\begin{eqnarray}
\label{eq:fitting}
f(L)=a+\frac{b}{L^c},
\end{eqnarray}
works pretty well, where $a$, $b$, and $c$ are fitting parameters.
We suspect that this is due to strong quantum fluctuations in the interaction
parameter regime (from $U/t=10$ to $16$) near the quantum phase transition.
The complex excitations would significantly change the finite size effect.
The finite size scalings are shown in Fig. ~\ref{fig:extrapolation}.
The extrapolated values of the AF and  VBS order parameters are plotted in Fig.~\ref{fig:phasediagram}.
Both of them drop to zero as the interaction parameter
approaches a small regime of $13<U/t<13.5$ from the opposite directions.
We have also checked the evolutions of total and kinetic energies.
Neither of them shows an obvious discontinuous behavior, which
suggests continuous transitions.

Certainly, there exist a few possibilities regarding to the nature
of these two ordering transitions: $a$) There is only a direct
2nd order phase transition as in the framework of the deconfined
criticality, i.e., the two orderings share the same critical value
of $U$; $b$) they exhibit two separate but very close 2nd order phase
transitions with a quantum disordered phase in between;
$c$) the same as in $b$) but with a small coexistence regime
of both orders; $d$) a weak 1st order transition between them.
To further address the nature of these transitions,
we perform the following scaling analysis.

\begin{figure} 
  \includegraphics[width=0.38\textwidth]{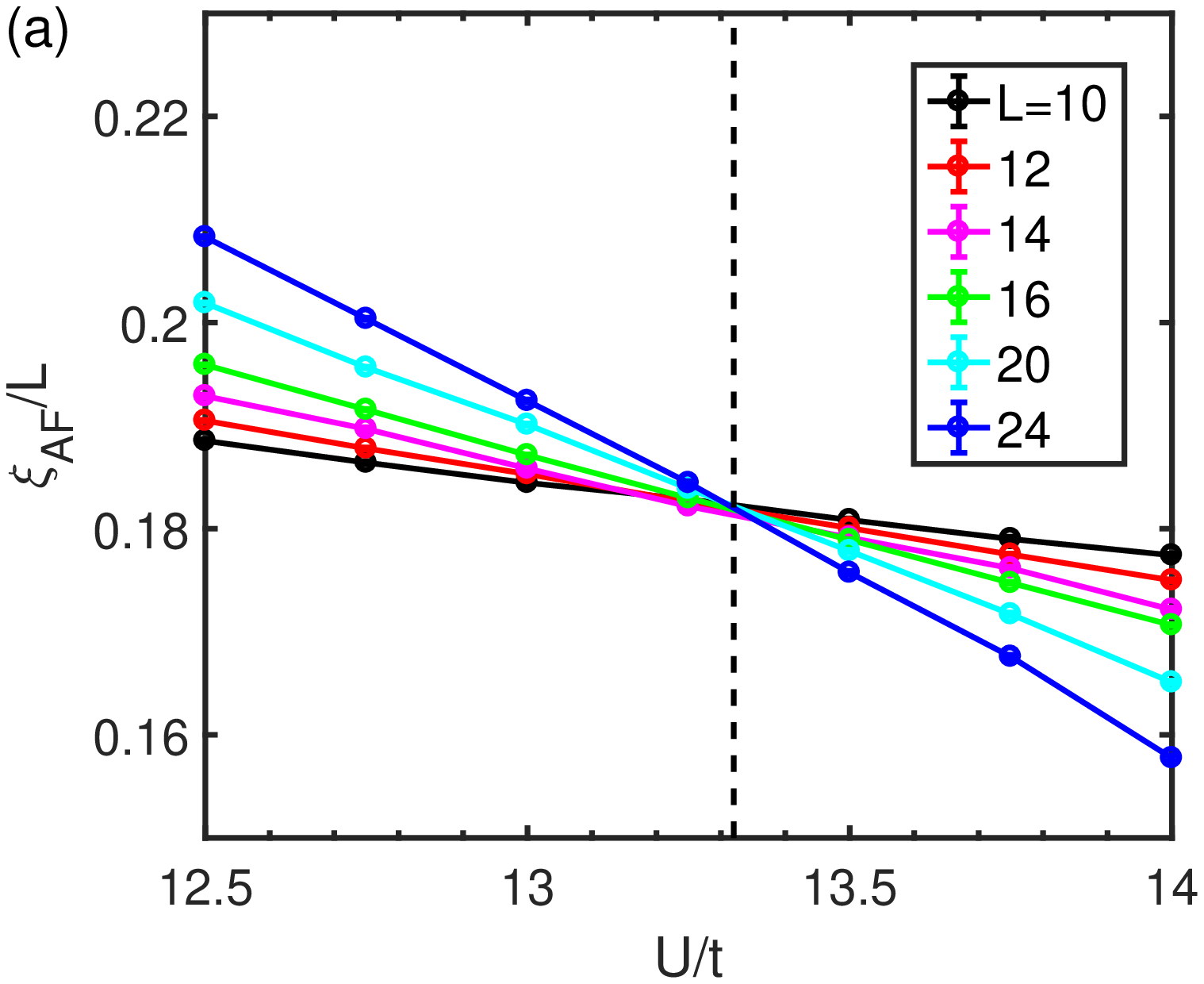}\\
  \includegraphics[width=0.38\textwidth]{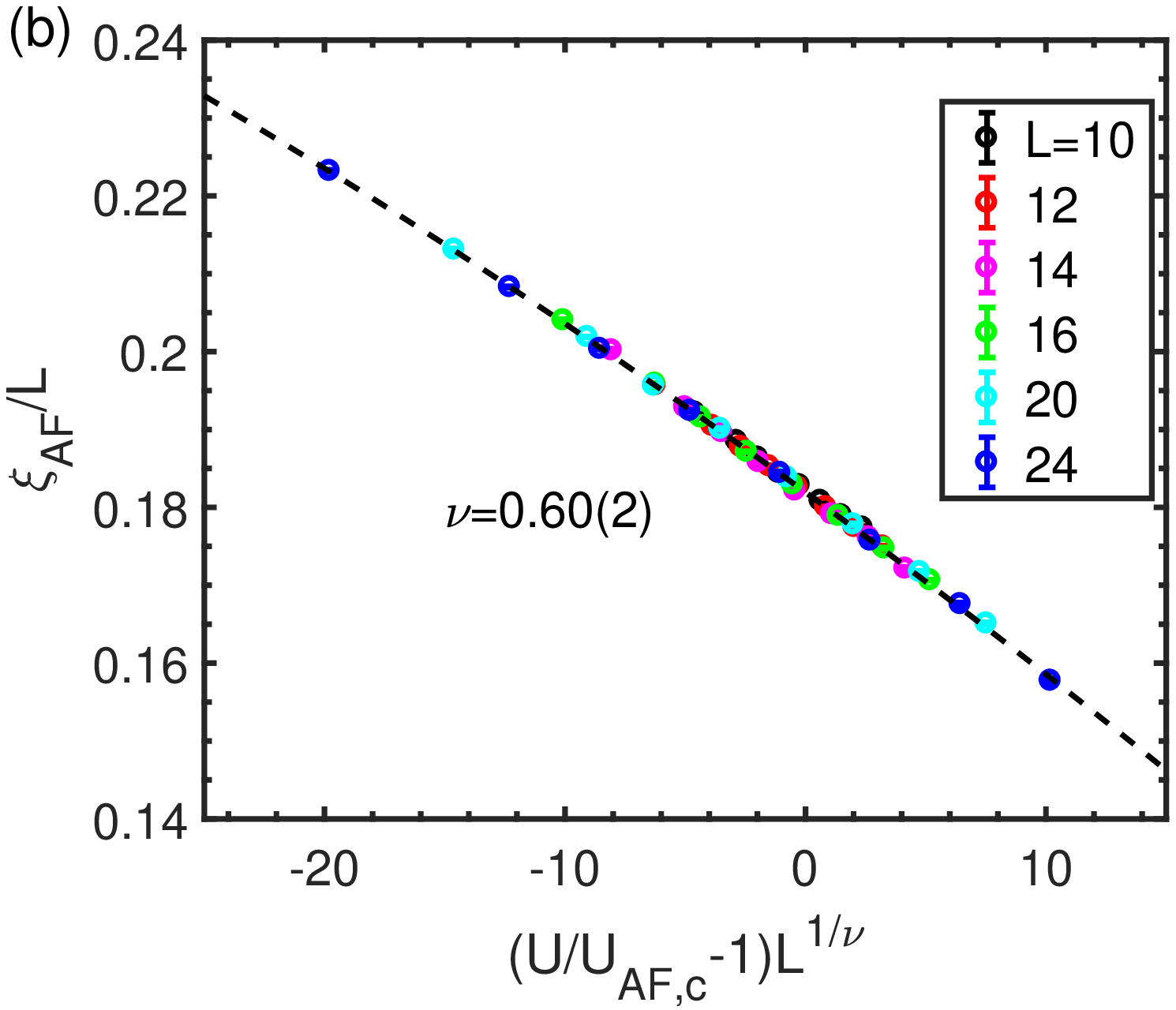}\\
  \includegraphics[width=0.38\textwidth]{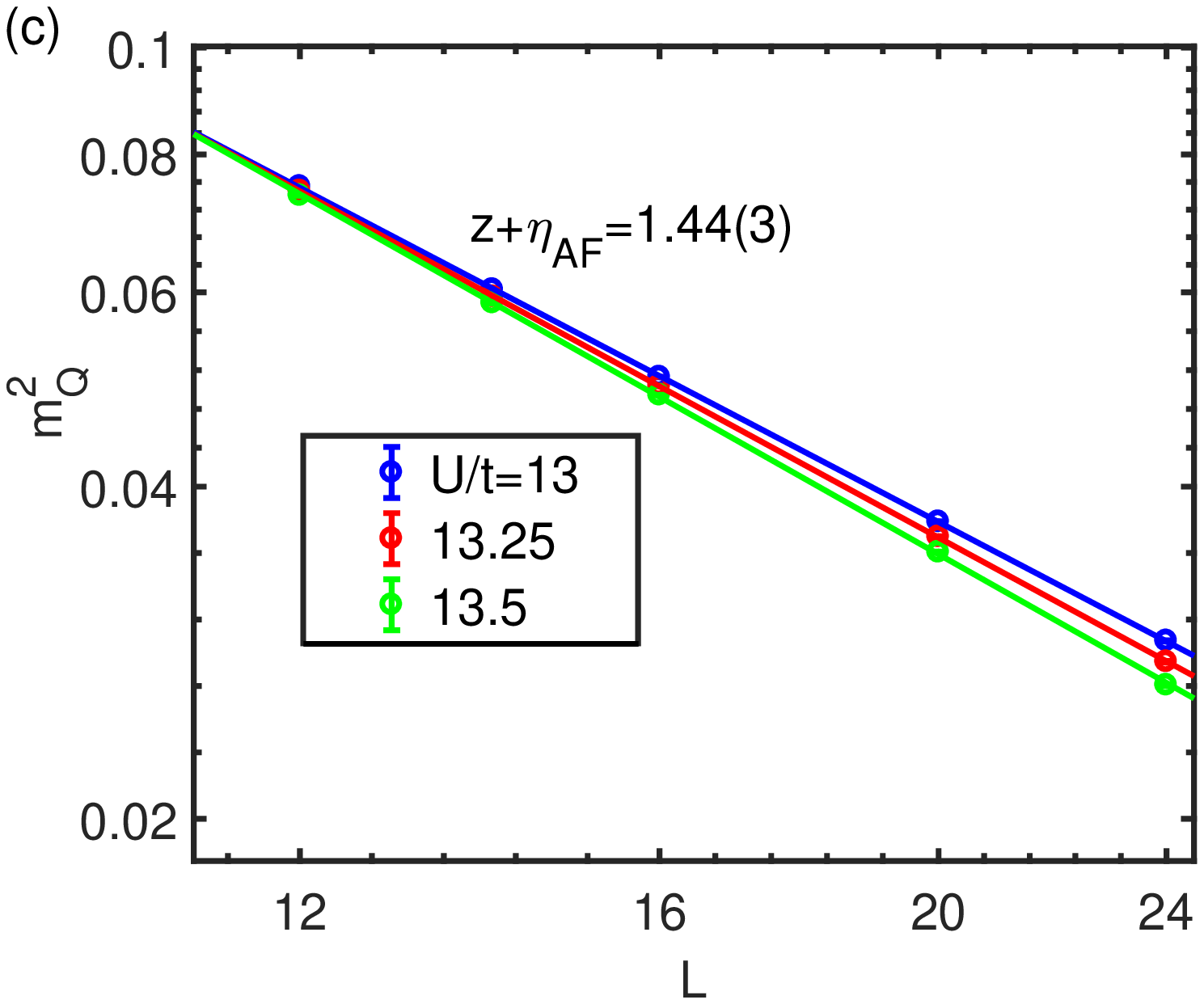}
  \caption{\label{fig:correlationlength_af}
    Scaling analysis of the AF correlation length $\xi_{\rm AF}$ and correlation $m_Q^2$.
    (a) $\xi_{\rm AF}/L$ versus $U$ shows a crossing point at $U_{\rm AF,c}/t=13.3\pm0.05$.
    (b)The data collapse of $\xi_{\rm AF}/L$ as a universal function of
    $(U-U_c)L^{1/\nu}$ with $\nu=0.60\pm 0.02$.
    (c) Log-log plots for $m_Q^2$ v.s. $L$ in the vicinity of $U_{\rm AF,c}$
    The fitting of the slopes gives rise to the
    anomalous dimension of $\eta_{\rm AF}$.
  }
\end{figure}

We first consider the scalings from the AF side.
The following definition of the AF correlation length is employed based
on spin-spin correlations
\cite{Sandvik_ACP_2010}
\begin{eqnarray}
\xi_{\rm AF}=\frac{1}{q}
\sqrt{\frac{m_Q^2 }{m_{Q+q}^2 }-1},
\end{eqnarray}
where $Q=(\pi,\pi)$ is the ordering wavevector, and $q$ is a small
deviation from $Q$ chosen as $q=(2\pi/L,0)$.
In Fig.~\ref{fig:correlationlength_af}($a$), we plot $\xi_{\rm AF}/L$ versus
$U$ at different values of $L$, and find that they cross at
$U_{\rm AF,c}/t=13.3\pm 0.05$, which is taken as the transition
point for the AF order.

Based on the critical value of $U_{\rm AF,c}$, we further perform the data
collapse as plotted in Fig.~\ref{fig:correlationlength_af}($b$)
according to the scaling function of $\xi_{\rm AF}$,
\begin{eqnarray}\label{eq:scalinghypothesis}
\xi_{\rm AF}(U,L)=L f\left[|U-U_{\rm AF,c}|L^{1/\nu_{\rm AF}}\right],
\end{eqnarray}
where the exponent of the divergence of correlation length
is determined to be $\nu_{\rm AF}=0.60\pm 0.02$.
Such an exquisite scaling behavior is a strong hint to a
continuous phase transition.

At a quantum critical point, the two-point correlation function in $d+1$ dimensions
is expected to be algebraic decay as
\begin{eqnarray}
\langle (-)^r m_r m_0\rangle &&\sim \int d^d q e^{i \vec q \cdot \vec r}
\int d \omega
\Big(\frac{1}{\omega^{2/z}+q^2}\Big)^{\frac{2-\eta_{\rm AF}}{2}} \nonumber \\
&& \sim \frac{1}{r^{d+z-(2-\eta_{\rm AF})}},
\end{eqnarray}
where $z$ is the dynamic critical exponent and $\eta_{\rm AF}$ is the anomalous
dimension \cite{Sondhi_RMP_1997}.
After the Fourier transformation, the structure factor at finite size $L$
scales as
\begin{eqnarray}
 m^2_Q  \sim  \frac{1}{L^{d+z-2+\eta_{\rm AF}}},
\end{eqnarray}
at large enough values of $L$.
In Fig.~\ref{fig:correlationlength_af} ($c$), $m^2_Q$ is plotted versus $L$ on a
log-log coordinate around $U_{\rm AF,c}$, which exhibits a good linear
behavior up to $L=24$.
From their slopes, $z+\eta_{\rm AF}=1.44\pm 0.03$ is found.
In our simulations, $z$ is difficult to determine accurately since it
requires the time evolutions of two particle Green's functions which,
however, tend to be gapless at the critical point.
If we adopt the $z=1$ directly following the prediction of the deconfined
critical theory\cite{Senthil_S_2004,*Senthil_PRB_2004,*Levin_PRB_2004},
we arrive at the anomalous dimension $\eta_{\rm AF}=0.44\pm 0.03$.

Based on the above analysis, the AF transition exhibits quite clear
evidence of a 2nd order phase transition.
Here we summarize the critical value of $U_{\rm AF,c}$ and
the two critical exponents for the AF transition as
\begin{eqnarray}
U_{\rm AF,c}&=& 13.3\pm 0.05, \nonumber \\
\nu_{\rm AF}&=&0.60\pm 0.02,  \ \ \,
\eta_{\rm AF}=0.44\pm 0.03.
\label{eq:anomalous}
\end{eqnarray}

\begin{figure}
\includegraphics[width=0.38\textwidth]{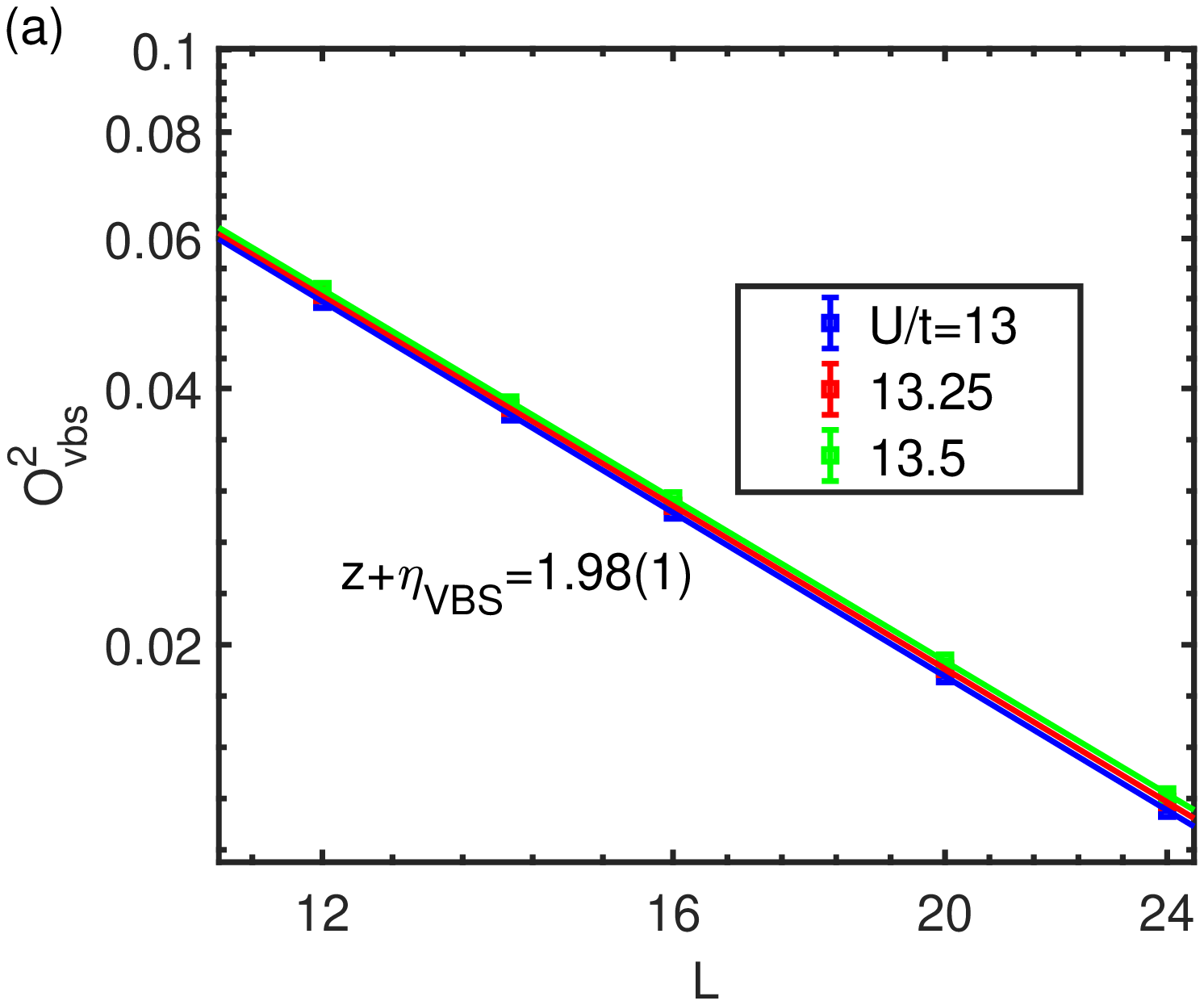}
\includegraphics[width=0.38\textwidth]{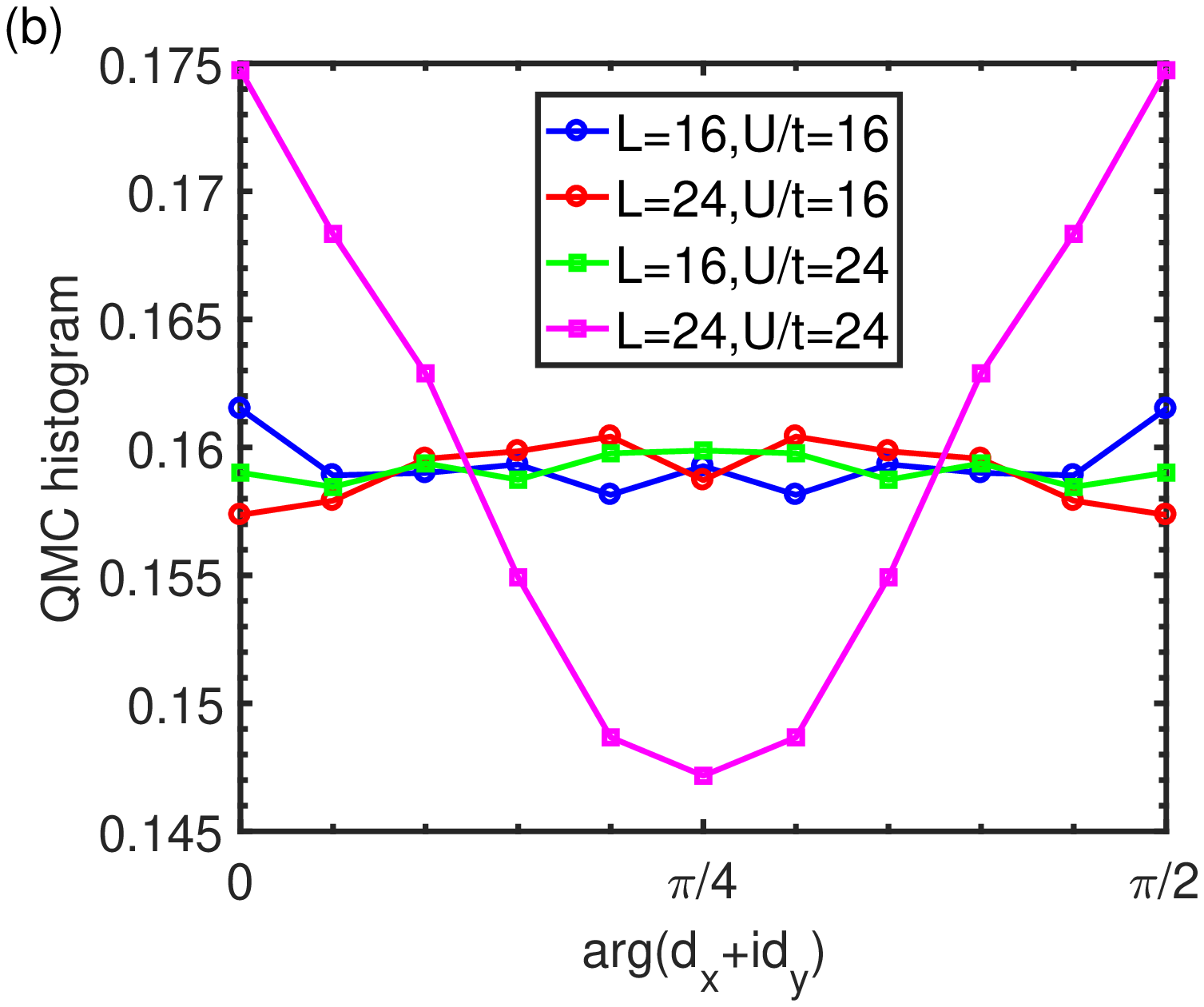}
\caption{\label{fig:vbs}
($a$) Log-log plots for $O_{vbs}^2$ v.s. $L$ in the vicinity of $U_{\rm AF,c}$.
The fitting of the slopes gives rise to the anomalous dimension $\eta_{vbs}$.
($b$) Histogram of the VBS configurations during the QMC simulations.
The result shows the VBS belongs to columnar type deep inside the VBS state.
}
\end{figure}

As for the correlations for the VBS orderings, unfortunately, it is 
difficult to obtain high quality data for the scaling of the VBS
correlation length to determine the critical value of $U_{\rm VBS,c}$
and compare it with $U_{\rm AF,c}$.
Nevertheless, since the VBS transition is very close to the AF one,
we present the log-log plot of $O^2_{vbs}(L) \sim (1/L)^{d+z-2+\eta_{\rm VBS}}$
for $U$ in the vicinity of $U_{\rm AF,c}$.
The fitting of the slopes yields $z+\eta_{\rm VBS}=1.98 \pm 0.01$, which
corresponds to
\begin{eqnarray}
\eta_{\rm VBS}=0.98 \pm 0.01.
\label{eq:anomalous_VBS}
\end{eqnarray}

The above anomalous dimensions $\eta_{\rm AF}$ and $\eta_{\rm VBS}$ are 
different from those obtained from the SU(6) $J$-$Q$ model \cite{Kaul_PRL_2012} indicating they indeed belong to different universality classes.
Our case is based on the fermionic SU(6) Hubbard model, and in its Mott
insulating state each site is in the self-conjugate representation, i.e.,
3 fermions per site, while the $J$-$Q$ model is equivalent to
the non-compact CP$^5$ model in which neighboring states belong
to the SU(6) fundamental and anti-fundamental representations.

In order to identify the type of the VBS order,
we plot the histograms of the VBS configurations during QMC simulations,
as shown in Fig.~\ref{fig:vbs}(b). Deep inside the VBS state 
(for large $U$ and large $L$),
the histogram shows larger weights at $\text{arg}(d_x+id_y)=0$ than $\pi/4$,
indicating that the VBS belongs to the columnar type.
However, near the phase boundary, the histograms are difficult
to tell which type of the VBS is, which in fact is consistent with 
an emergent $U(1)$ symmetry in the framework of deconfined criticality.
\cite{Sandvik_PRL_2007}

\section{summary and discussion}
In summary, we have performed a large scale projector QMC simulations on
the half-filled SU(6) Hubbard model in the square lattice.
As $U$ increases, we have found a crossover at $U^*/t\approx9$
from the Slater-AF to the Mott-AF insulators.
As $U$ further increases, a (signature of)
continuous phase transition at $U_c/t=13.3\pm 0.05$ from the
Mott-AF to Mott-VBS states.

Several remarks of these numerical observations are given as follows:
(1)The finite size extrapolations in this work are based on the power
law fitting in Eq.~\ref{eq:fitting}, which is different from most studies,
especially the SU(2) Heisenberg model where the cubic-order polynomial
works very well \cite{Neuberger_PRB_1989,*Sandvik_PRB_1997}.
The difference may be rooted in the stronger quantum fluctuations
from the higher symmetry group $SU(N)$.
A full understanding requires more sophisticated knowledge of the excitation
properties of the SU(N) Hubbard which is left in future studies.
(2) We have also tried to obtain the universal plots of
$\xi_{\rm VBS}/L$ and $L^{z+\eta_{\rm VBS}}O_{vbs}^2$
versus $|U-U_c|L^{1/\nu_{\rm VBS}}$ but failed.
Of course, this may be caused by insufficient lattice sizes up to $L=24$
in our simulations.
However, another possible reason may be the recently proposed two-length
scaling hypothesis \cite{Shao_S_2016}, which requires very large
lattice sizes difficult to reach for the determinant QMC simulations
for fermions.
(3) By symmetry analysis, the observed AF-VBS phase transition belongs to a
broader universality class governed by a $U(N)/[U(m)\otimes U(N-m)]$
nonlinear sigma model beyond the CP$^{N-1}$ model corresponding to $m=1$.
The full understanding, e.g. critical exponents, of the $m>1$ models
calls for more elaborated theoretical efforts via e.g.
$1/N$ expansion\cite{Wang_a_2018}, or,
renormalization group analysis\cite{Das_2018} in the future.

\section{acknowledgement}
D. W. is supported by NSFC under grant No. 11874205.
L. W. is supported by supported by the Ministry of Science and Technology of China 
under the Grant No. 2016YFA0302400 and the Strategic Priority Research Program of 
Chinese Academy of Sciences Grant No. XDB28000000.
C. W. is supported by AFOSR FA9550-14-1-0168.
The numerical calculations were performed on Tianhe-II platform 
at the National Supercomputer Center in Guangzhou.

\bibliography{su6dqcp}
\end{document}